\documentclass[conference]{IEEEtran}

\usepackage{setspace}
\usepackage{algorithm}
\usepackage{algorithmic}
\usepackage{amssymb}
\usepackage[figuresright]{rotating}
\usepackage{graphicx}
\usepackage{amssymb}
\usepackage{epstopdf}		
\usepackage{caption}
\usepackage{xcolor}

\usepackage{subcaption}

\ifCLASSINFOpdf
\else
\fi

\usepackage{amsmath,amsfonts,amsthm} 

\usepackage{graphicx}
\usepackage{epstopdf}

\usepackage{float}

\newcommand{\titolo}{Building accurate HAV exploiting User Profiling and Sentiment Analysis}

\begin{document}
%
\title{\titolo}

\author
{
\IEEEauthorblockN{Alan Ferrari, and Angelo Consoli}
\IEEEauthorblockA{\\Information Systems and Networking Institute \\ University of Applied Sciences\\ of Southern Switzerland (SUPSI)\\
Manno, Switzerland\\
firstName.lastName@supsi.ch}
}

\maketitle

\begin{abstract}
Social Engineering (SE) is one of the most dangerous aspect an attacker can use against a given entity (private citizen, industry, government, ...). In order to perform SE attacks, it is necessary to collect as much information as possible about the target (or victim(s)). The  aim of this paper is to report the details of an activity which took to the development of an automatic tool that extracts, categorizes and summarizes the target interests, thus possible weaknesses with respect to specific topics. Data is collected from the user?s activity on social networks, parsed and analyzed using text mining techniques. The main contribution of the proposed tool consists in delivering some reports that allow the citizen, institutions as well as private bodies the screening of their "exposure" to SE attacks, with a strong awareness potential that will be reflected in a decrease of the risks and a good opportunity to save money.
 \end{abstract}

\IEEEpeerreviewmaketitle

\section{Introduction}
Nowadays information are one of the most sensible and valuable property that a given organization owns thus if they are lost or, even worst, stolen their business can face severe consequences. 

One of the core techniques to gain unauthorised access to sensible information is Social engineering that can be defined as the psychological manipulation of users to gain access to sensible information. Social engineering is one fo the most powerful tool in the hand of every attacker.

An hacker, that is interested to perform  such kind of attacks must following a precise path; the first action consist in collecting information about the potential target and then use them to build a credibile hook. Only after the two previous steps the attacker can finally perform the attack. Information about a given victim can be collected in a lots of different ways (e.g. shoulder surfers, ...) and all of them form the so called Human Attack Vector (HAV).

Probably the most used methodology to gather victim's information consists in analyzing their online presence. In Social Networks users tend share a lot of personal information (e.g. politics ideas, holidays, ...) that an attacker can use to build the HAV. 

 The challenge every attacker has to face  is the enormous amount of information users tend to publish inside social network, for instance every second more than 6000 tweets are published\footnote{http://www.internetlivestats.com/twitter-statistics/}  thus the analysis might requires quite a long time.
 
The \textbf{key goal} of the following work is to illustrate how an attacker can parse a large amount of user's information automatically thus build the HAV in a very short time and with very little effort.  To reach the objective \textbf{Text-mining}, a branch of artificial intelligente that has as main goal the extraction of information from unstructured text, is used.

Victim's opinion is crucial to perform sensible and credibile attacks; to gather those information   \textbf{opinion mining}  or \textbf{sentiment analysis } is used, those techniques lay in the field of text mining and consist in the extraction of the polarity from a given sentences.

\section{Related Work}
Social Engineering can be describe as the art of influencing people to obtain sensible information, a Social Engineer manipulates the victim and convince her to divulgate confidential information.  Cognitive bias, that can be described as a specific attribute of human decision-making process, is at the base of those techniques. Following the category proposed in \cite{krombholz2015advanced}, Social Engineering approaches can be divided as:

\begin {table}[t]
\begin{center}
  \begin{tabular}{| l | r |   }
    \hline
    Name & Size [\# Users]  \\
\hline
Google+	&1'600'000'000  \\
Facebook	& 1'280'000'000  \\
Twitter	& 645'750'000  \\ 
Qzone	& 48'000'0000  \\
Sina Weibo &	30'000'0000  \\
Instagram	 & 3'0000'0000  \\
Habbo	& 26'800'0000  \\
VK	& 249'409'900  \\
Tumblr	& 226'950'000  \\
LinkedIn & 	200'000'000  \\
Renren	& 160'000'000  \\
Bebo	 & 117'000'000  \\
Tagged	& 100'000'000  \\
Orkut	& 100'000'000  \\
Netlog	& 95'000'000  \\
Friendster	 & 90'000'000  \\
hi5 & 	80'000'000  \\
Flixster	 & 63'000'000  \\
MyLife &	51'000'000  \\
Classmates.com &	50'000'000  \\
Sonico.com	& 50'000'000  \\
Plaxo	& 50'000'000  \\
douban	& 46'850'000  \\
Odnoklassniki	& 45'000'000  \\
Viadeo	& 35'000'000  \\
Flickr	& 32'000'000  \\
  \hline    
  \end{tabular}
  \caption{List of most populated social network websites ordered by the number of users (Source: http://www.wikipedia.com on September 2016.}
  \label{tab:fr}
\end{center}
\end {table}

\begin{itemize}
\item \textit{Physical Approaches: } where an attacker perform a physical action to collect victim's information. The most common method is \textit{Dumpster Diving} \cite{granger2001social} where an attacker search inside victim's garbage some useful information.
\item \textit{Social Approachers: } where an attacker uses socio-psycological techniques to manipulate the victim. Persuasion is one of the most used methods that can i.e. use Authorities as a main tool  to convince the user, however other techniques can be used as well i.e. user's curiosity can be used as well . According to  \cite{granger2001social} the majority of social approaches are done by phone.
\item \textit{Reverse Social Engineering: }  according to \cite{nelson2001methods} reverse social engineering consist in create a situation where the victim contacts the attacker, for instance sabotaging a company's computer and pretending to be a technician.
\item \textit{Technical Approaches}: attackers uses the online presence of the users to collect information about them; social network are the most common place where a user put informatioan bout herself. 
\item \textit{Social-technical Approaches: } the combination of Social approaches and Technical approaches have create one of the most powerful approaches an attackers cant use. An example can be to put a trojan-horse malware in a USB key or folder and call it with an appealing name, thus exploiting curiosity. 
\item \textit{Office Communication: } where an attacker uses traditional communication tool in an office environment (e-mail, ...) to perform an attack; such tools add credibility to attacker. 
\item \textit{Extenal Communication: } every organisation has a set of external partners (i.e. web/e-mail hosting, ...) the key idea here is that an attacker let the victim believes she is one of the external and then gain the victim's confidence. 
\end{itemize}
The focus of this paper is in the Socio-Technical approach and social network are the key tool used to gather user's sensible information and build the HAV. The first notable work on this are has been presented by Huber et. al. \cite{huber2009towards} where authors uses social network as tool to gather vicim's information and perform an attack. The work already mentioned has some sever limitations because the framework   collects only structured information (e.g. victims contacts and profile). The goal of this work is to move one step further and used current state of the art in text mining to parse and collection information from raw data that a user upload in their social network profile (i.e. user's posts) and build accurate HAV.

\section{System Architecture}
This section provides a deep description of our system and the key challenges have been faced in order to built it.

\begin{figure*}[t]
\centering
\includegraphics[scale=.4]{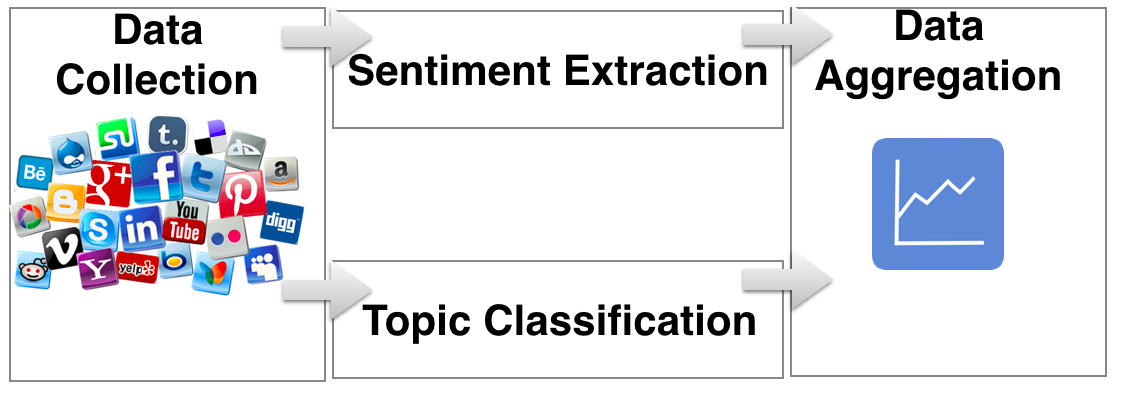}
%
\caption{Architecture of the System  that is a set of structure layers from information gathering to aggregate information visualization. }
\label{fig:rarc}
\end{figure*}

Figure \ref{fig:rarc} shows the rationale of the system architecture; once the system is executed it takes as input the social profile of a given users, with those information it downloads it and extract two kind of information; the topics that a user addresses and the user sentiment expressed in the text. After, topics are opinion are aggregated and showed to the  attacker, at this point he has a clear picture of the information that he might need to build the HAV.

\subsection{Information Collection}
With the advent of Web2.0  users have gained the great opportunity to express their feelings and ideas and share all of them with the worlds thanks to tools like blogs, forums, ...).  During past years  Social network have become the major provider of user personal information. 

Following \cite{ellison2007social}  a Social Network can be defined as  a service that allows its users to create a profile (with a certain degree of visibility) where they can connect to their friends and share the content they create in different forms (from text to vides).

The key element is that users tend to include social network in their daily live including a lot of personal information that range from current location and activity to political view or preferred food. 

 Publicly available Software API that the majority social network offers to access their data programmatically are used to collect social-information .  Social Networks are a global phenomena they they are used by a tremendous number of users (see Table \cite{ab:fr} for reference); in our word we adopt publicly available API  that are offered by:
 
\begin{itemize}
\item \textbf{Facebook} \footnote{http://www.facebook.colm}: that is social network created by Mark Zukemberg in 1994 wich give servers to a tremendous set of users and a multitude of platforms. Within Fadebook uses can share a lot of content ranging from text posts to images and even videos.  
\item \textbf{Twitter} \footnote{http://www.twitter.com}: created in 2006 twitter is a social network that has the great peculiarity to maintain the user input as short and simple, in fact they are allowed only to 'publish the so called tweet that is a 140-characters long message.   Twitter it is widely adopted reaching a tremendous amount of users that made an even bigger number of tweet per day.
\item \textbf{Linkedin} \footnote{http://www.linkedin.com}: created in 2002, LinkedIn is a social network with main target to connect users  that share a professional relationship. Within LinkedIn users can share their professional profile and find or post new jobs announce or any information that is pertinent to their professional background in the form of posts. 
\item \textbf{G+} \footnote{http://www.g+.com}: Created in 2013 by Goole it is  defined as an interest-based social network offer the same functionalities we found in the previously mentioned  social network but with the concept of circle that allows users to organize their contact  in groups or list for sharing.
\end{itemize}

In the implementation we limit  the platforms to \textbf{Twitter} and \textbf{G+} because they are the only two platforms that allows developers to access the content of a single users and offer a dedicated API to to that job; even thought it is possibile to gain the same access to other Social Network  at the moment we are writing this paper this functionality is blocked and strictly forbidden by the social network usage rules. 

\subsection{Sentiment Extraction}
We can define the Sentiment Extraction process as the usage of sentiment analysis algorithms to each user text. In general Sentiment Analysis or Opinion mining extract the attitude or the polarity of a given text. In our implementation we use state of art algorithm based on Deep Learning strategy. In \cite{socher2013recursive} authors propose a dedicated deep learning architecture based on  the concept of \textit{Recursive Neural Tensor Network} for Opining Mining. Authors also trained the network with 11'855 sentences obtaining a boost in accuracy compared to previously mentioned solutions.  The network returns a classification between very negative to very positive of a given text. 

In our work for each given user's text we compute and store the corresponding sentiment.

\begin{algorithm}
\caption{W2V Distance}
\label{alg:alg1}
\begin{algorithmic}
\STATE  \textbf{Input:} $t$ (topic) $K$ (text)
\STATE $r = \infty$
\FOR{ \textbf{all} $k \in K$}
\FOR{ \textbf{all} $w \in c_t$}
\STATE $r' =  d(k, w)'$  (d is the Euclidean distance)
\IF{$r > r'$}
\STATE $r = r'$
\ENDIF 
\ENDFOR
\ENDFOR
\STATE \textbf{Return} $r$
\end{algorithmic}
\end{algorithm}

\subsection{Topic Classification}

Each text (tweets or post) a given user upload in the Social Network must be categorized into a topic such that the attacker has access to aggregate information that are easy and fast to given an interpretation.  In the framework an attacker must  provide a list of topic ($T$) in conjunction with a set of texts that describe it. An attacker is free to insert watherver category he needs, for instance if an attacker is interested to know the technogloy-awareness of a given user he can created the category "Computer" and train the system with the Wikipedia pages connected to that topic.

To create a dynamic classification system K-Means cluster solution is used.

\subsubsection{K-Means Variant}

K-Means cluster algorithm is a popular techniques used in data mining to perform clusterization, in other word to classify a given input into a pre-defined cluster or group.
We define the traditional k-means as following: assume you have a  finite set of elements $n \in N$  and $c  \in C$ that is a cluster. Each cluster is defined by its centroid that is computed by a function  $d$ of all the $c \in c$. If $n' \in N$ is passed as input in the system it is associated by the $c$  where:
 \[
 d(n', c) <  d(n', d) \forall  (c,d) \in C
\]
Even thought K-means has been created for numerical domains it has been applied to Text Mining  as well. Traditionally researchers converts words in numbers using a dictionary based search,  thus the centroid is formed by the words that are part of a given topic. Assume someone wants to classify a document it uses as distance function the number of words that stay in both the document and the cluster. 

The already mentioned approach has provided good results, however in the following paper we propose an extension that uses state of the art in text-ming to define a better distance function.

The starting point is a model named Word2Vec \cite{goldberg2014word2vec} provided by Goldberg et. al. The key output of Word2Vec algorithm is to produce word embedding in other word to transform a word into a meaningful numerical representation. Word2Vec uses a two-layer neural network  that produces output in a high dimensional space. If properly trained,  the great advance in using such kind of algorithms is that the distance between two vectors generated by two distinct words (in our case computed by the Euclidean distance) icontains information about the semantic-distance between them. We use a pre-trained vector Word2Vec that has been trained with Google News Dataset (more than 3 billion words)\footnote{https://code.google.com/archive/p/word2vec/}.

We define our algorithm as follow: for any given topic $t \in T$ we define the Cluster Centroid $c_t$ as the the coordinate of the nouns that represents such topic, we extract them from the input documents.  

To classify a input text $K$ we extract each \textbf{nouns} in the text  and we computed the minimum distance between the nouns and $c_t$ defined in the Algorithm \ref{alg:alg1}

The sum of all the distacene between the words in $K$ and the cluster is defined as function  $l(K, c_t)$. We classify $K$ un a given cluster $r$according to the following equation:

\[
 r = t | l(K, c_t) < (K, c_i) \forall t,i \in T
\]
Where $r$ is the corresponding topic, for word the system also stores the word frequency that is a usesful metrics to  have an idea about the intensity a given users talk about a given topic.
\section{Visualization}
At the end the system allows user to visualize each topic and  the connect text (classified with algorithm illustrated before) with the  connected sentiment (computed with the algorithm illustrated before). Therefore, for each topic we also provide a aggregated information by showing the most frequent sentiment per topic; we argue that this information will be used to the attacker to create better and credibile hook for her attacks.

\section{Conclusion and Future Work}
In this work we have presented the architecture of our framework that allows users to download and automatically extract sensible information from unstructured raw data that potential victims put inside social networks. Each information are classified into a topic and the user's opinion about that topic is computed as well.

We plan to use the results of our work in build an advance tool to sensibilize a given organization agains Social Engineering attacks allowing it to save money and reduce risks that such kind of attacks can generate..

\bibliographystyle{unsrt}	
\bibliography{se-opiniongminig}

\end{document}